\documentclass[twocolumn]{aastex631}

\begin{document}

\title{Constraining the clustering and 21-cm signature of radio galaxies at cosmic dawn}

\correspondingauthor{Sudipta Sikder}
\email{sudiptas@mail.tau.ac.il}

\author{Sudipta Sikder}
\affil{School of Physics and Astronomy, Tel-Aviv University, Tel-Aviv, 69978, Israel}

\author{Rennan Barkana}
\affiliation{School of Physics and Astronomy, Tel-Aviv University, Tel-Aviv, 69978, Israel}

\author{Anastasia Fialkov}
\affiliation{Institute of Astronomy, University of Cambridge, Madingley Road, Cambridge, CB3 0HA, UK}

\affiliation{Kavli Institute for Cosmology, Madingley Road, Cambridge, CB3 0HA, UK}



\begin{abstract}
The efficiency of radio emission is an important unknown parameter of early galaxies at cosmic dawn, as models with high efficiency have been shown to modify the cosmological 21-cm signal substantially, deepening the absorption trough and boosting the 21-cm power spectrum. Such models have been previously directly constrained by the overall extragalactic radio background as observed by ARCADE-2 and LWA-1. In this work, we constrain the clustering of high redshift radio sources by utilizing the observed upper limits on arcminute-scale anisotropy from the VLA at 4.9~GHz and ATCA at 8.7~GHz. Using a semi-numerical simulation of a plausible astrophysical model for illustration, we show that the clustering constraints on the radio efficiency are much stronger than those from the overall background intensity, by a factor that varies from 12 at redshift 7 to 30 at redshift 22. As a result, the predicted maximum depth of the global 21-cm signal is lowered by a factor of 5 (to 1700~mK), and the maximum 21-cm power spectrum peak at cosmic dawn is lowered by a factor of 24 (to $2\times 10^5$~mK$^2$). We conclude that the observed clustering is the strongest current direct constraint on such models, but strong early radio emission from galaxies remains viable for producing a strongly enhanced 21-cm signal from cosmic dawn.

\end{abstract}



\section{Introduction} \label{sec:intro}

The Absolute Radiometer for Cosmology, Astrophysics, and Diffuse Emission 2 (ARCADE-2) measurements of the absolute sky brightness at GHz frequencies provides evidence for a strong radio background that is consistent with the cosmic microwave background (CMB) radiation at higher frequencies. However, it significantly diverges from the blackbody spectrum at low frequency, as demonstrated by \citet{fixsen11}. The result of ARCADE-2 was confirmed using the first station of the Long Wavelength Array (LWA-1)  \citep{dowell18} in the frequency range 40-80 MHz. Since the amplitude of this radiation is considerably larger than anticipated based on observed radio counts of known galactic and extragalactic sources \citep{Singal2018}, the origin of this radio background is an interesting astrophysical mystery. While extragalactic explanations for the origins of this radio background suffer from various challenges
\citep{Singal2010, Kogut2011a, Vernstrom2011, Vernstrom2014, Condon2012}, previous works showed that the excess radio background could be produced by early radio loud quasars \citep{Bolgar2018}, active galactic nuclei (AGN) from cosmic dawn \citep{EwallWice2014}, the first generation of supermassive black holes \citep{Biermann2014}, or high redshift star-forming galaxies \citep{Condon1992}. It is important to remember that the Galactic contribution is uncertain \citep{Subrahmanyan:2013}, but the observed radio background certainly puts an upper limit on the contribution from any extragalactic source population. 

After the reported EDGES measurement of an excess 21-cm absorption signal \citep{bowman18}, two classes of explanations were immediately offered for the anomalous EDGES absorption trough: an excess cooling of the ordinary matter through scattering with dark matter \citep{Barkana2018}, and the enhancement of the CMB by an excess radio background as explored in various works \citep{bowman18, feng18, fialkov19, mirocha19, ewall20, Mebane2020, Sikder2023}. More recently, the SARAS-3 experiment \citep{SARAS3} found a discrepancy between the EDGES absorption profile and their own measurements, with a confidence level of $95\%$. Along with forthcoming findings from EDGES and SARAS, upcoming global signal experiments such as PRIZM \citep{prizm2019}, SCI-HI \citep{scihi2014}, REACH \citep{reach2022}, and MIST \citep{mist2023} are poised to offer further insight into the observational quest for the global 21-cm signal.

The tentative EDGES signal, along with the extragalactic radio background measured by ARCADE-2, have served as a driving force for exploring astrophysically-grounded excess radio models and analyzing their impact on the 21-cm signal. \citet{Reis2020} first incorporated a non-uniform radio excess from high redshift galaxies into a 21-cm semi-numerical code and investigated its effect on the global signal as well as on the 21-cm power spectrum. In our recent work \citep{Sikder2023}, we showed that a more accurate calculation of this radio excess, fully including the line-of-sight effect of individual radio galaxies, enhances the 21-cm power spectrum by up to two orders of magnitude during cosmic dawn (depending on the astrophysical parameters). In these papers, we considered the need not to overproduce the observed radio background, and showed that the resulting constraint on the radio efficiency is strongly redshift-dependent, so that the efficiency could have been quite high at early times. 

In this \textit{Letter}, we consider observational constraints not just on the overall intensity of the cosmic radio background (CRB), but also on its clustering. The most stringent constraints come from radio observations with the Very Large Array (VLA) at $4.9$ GHz \citep{Fomalont1988} and the Australia Telescope Compact Array (ATCA) at $8.7$ GHz \citep{Subrahmanyan2000}. \citet{holder2014} pointed out that clustering is a strong constraint on low-redshift radio sources, and we consider much higher redshifts. Our semi-numerical simulations of the early Universe allow us to predict the contribution of high-redshift galaxies to the observed clustering. Although the CMB dominates at these relatively high frequencies, even a population that makes a sub-dominant contribution can exceed the observed limits if that population is sufficiently strongly clustered. This can happen at high redshifts, when bright radio galaxies were rare and each served as a background source for strong 21-cm absorption. Using the existing upper limits, we aim to constrain the high redshift radio clustering and investigate the consequences for the possible range of the cosmic dawn 21-cm signal. 

\section{Methodology}

\subsection{Astrophysical model}

To calculate the mean brightness temperature of the excess radio background at the redshifted wavelength of 21-cm radiation as well as the 21-cm signal itself, we rely on our simulation code that we have named 21cmSPACE (21-cm Semi-numerical Predictions Across Cosmological Epochs), as detailed in previous works \citep[see e.g.,][]{visbal12, fialkov14, cohen17, fialkov19}. The simulation yields the 21-cm brightness temperature within a cosmological (comoving) volume of $[384\,{\rm Mpc}]^3$, at a resolution of 3 comoving Mpc, spanning a wide range of redshifts from 6 to 35. We focus on models with extra-strong radio emission from early galaxies, including the line-of-sight effect of individual radio galaxies \citep{Reis2020,Sikder2023}. 

The basic simulated model has seven astrophysical parameters: star formation efficiency ($f_{\star}$), minimum mass of star forming halos (given by a minimum circular velocity $V_c$, except that various types of feedback are also included), X-ray efficiency ($f_{\rm{X}}$, where unity corresponds to the typical observed value for low metallicity galaxies), X-ray SED parameters (power-law slope $\alpha$ and minimum X-ray energy $E_{\rm{min}}$), and reionization parameters (ionization efficiency $\zeta$ and maximum mean free path of ionizing photons $R_{\rm{mfp}}$). In addition, for models with strong radio emission we follow our previous papers and assume, based on the empirical relation of \citet{gurkan18}, that the radio luminosity per unit frequency of a galaxy is proportional to its star formation rate. We also assume a radio spectral slope that corresponds to synchrotron radiation and roughly agrees with the slope of the observed extragalactic background. An additional parameter is introduced here, i.e., $f_{\rm{Radio}}$, the normalization of the radio emissivity relative to the star formation rate, where $f_{\rm{Radio}} = 1$ corresponds to present-day star-forming galaxies. 

\subsection{Simulation output}

In order to obtain the radio observables from the simulation, we first note that the observed 21-cm brightness temperature relative to the CMB (in a pixel in some direction at redshift $z$) is:
\begin{equation}
    T_{21}^{\rm full} = \frac {\left( T_{R,\rm los} + T_{\rm CMB} \right) e^{- \tau_{21}} + T_S \left( 1 - e^{- \tau_{21}} \right) - T_{\rm CMB}} {1+z}\ .
    \label{eqn:T21_full}
\end{equation}
Here $T_{\rm{R, los}}$ is the brightness temperature of the radio background from sources lying behind the pixel along our line of sight, $T_S$ is the spin temperature and $\tau_{21}$ is the 21-cm optical depth, where the latter two quantities depend on $T_{\rm{Radio}}$, the isotropically-averaged radio intensity at the relevant pixel \citep[for details, see][]{Sikder2023}. 

Consider the $\tau_{21} = 0$ limit of this expression:
\begin{equation}
    T^{\tau = 0}_{21}(z)  =  \frac{T_{R,\rm los}}{1+z} \ .
    \label{eqn:T21_Radio}
\end{equation}
When there is no significant radio background (i.e., $T_{R,\rm los} \ll T_{\rm CMB}$), this $\tau_{21} = 0$ term vanishes. Even when there is a radio background, the $\tau_{21} = 0$ term is normally subtracted (e.g., in our previous papers), since it represents a signal component that is independent of 21-cm emission or absorption (as it does not depend on $\tau_{21}$), and has a spectrum that is similar to the Galactic synchrotron foreground. In 21-cm observations, this component cannot be distinguished from Galactic synchrotron, and is thus automatically removed when foreground removal is applied (under the assumption of a smooth power-law synchrotron spectrum), whether we are considering radio interferometers or global experiments. Here, however, we are interested in the total radio background, both its mean level and its clustering. The $\tau_{21} = 0$ term strongly dominates the overall background, since $\tau_{21}$ is typically at the percent level. 

Thus, we separate the 21-cm signal into two components. For the excess radio background at the redshifted wavelength of 21-cm radiation we adopt the $\tau_{21} = 0$ term of eq.~(\ref{eqn:T21_Radio}). On the other hand, when we consider standard 21-cm observations, the relevant (foreground subtracted) 21-cm brightness temperature is found by subtracting eq.~(\ref{eqn:T21_Radio}) from eq.~(\ref{eqn:T21_full}), which yields \citep{Sikder2023}
\begin{equation}
T_{21}(z) = \frac { 
 T_S - \left( T_{R,\rm los} + T_{\rm CMB} \right) } {1+z}\ \left( 1 - e^{- \tau_{21}} \right)\ .
 \label{eqn:T21new}
\end{equation}

\subsection{Mean radio background} 

Consider a particular astrophysical model, with all the parameters of the high-redshift galaxies held fixed except that $f_{\rm{Radio}}$ is allowed to vary (and the total radio emission varies linearly with it). 
Based on the reported measurements of the extragalactic radio background at frequencies of 3, 8, and 10 GHz by ARCADE-2, the maximum possible radio background from unresolved sources can be written as follows \citep{holder2014}:
\begin{equation}
\langle T^{f_{\rm{Radio}}=1}_{21}(z)\rangle  f^{\rm{mean}}_{\rm{Radio}}(z) = 
T_{\rm{arcade}} - T_{\rm{counts}} \ .
\label{eqn:fRadio_mean}
\end{equation}
Here, on the right-hand side $T_{\rm{arcade}} = 28.3 \  \rm{K} \left[\frac{\nu(z)}{310}\right]^{-2.6}$ is the observed extragalactic radio temperature (after adding twice the error in order to get a $2\sigma$ upper limit), $T_{\rm{counts}} = 0.23 \  \rm{K} \left[\frac{\nu(z)}{1000}\right]^{-2.7}$ is the expected temperature of the radio sky, based on extrapolating known source counts \citep{holder2014,gervasi2008}, and $\nu(z) = 1420~\rm{MHz}/(1+z)$. On the left-hand side, $\langle T^{f_{\rm{Radio}}=1}_{21}(z)\rangle$ is the mean (volume averaged) brightness temperature from the radio background at the redshifted wavelength of 21-cm radiation, $T^{\tau = 0}_{21}$, in a simulation with $f_{\rm{Radio}}=1$. Thus, $f^{\rm{mean}}_{\rm{Radio}}$ is the highest value of $f_{\rm{Radio}}$ (at each redshift) that is consistent with the observed upper limit on the mean extragalactic radio background.

\subsection{Angular power spectrum}

To calculate the angular power spectrum of the excess radio background at the redshifted wavelength of 21-cm radiation from our model, we use eq.~\ref{eqn:T21_Radio} (since the $\tau_{21} = 0$ term dominates). Given the 3D brightness temperature cube at a redshift $z$, we choose a random slice to calculate the two-dimensional power spectrum of radio fluctuations, $P_{\rm{2D}}(k)$. We then use the power spectrum to find the squared fluctuation
\begin{equation}
    \Delta^2(l) = \frac{1}{2\pi}\left(\frac{l}{R}\right)^2P_{\rm{2D}}(l/R) \ [\rm{mK^2}]\ ,
\end{equation}
where the multipole number $l$ is given by $l = kR$ and $R$ is the distance (in Mpc) corresponding to the redshift $z$ (where both $k$ and $R$ are comoving). 
Finally, we average this over all the slices of a particular simulation box at $z$, and calculate the dimensionless (fractional) angular fluctuation  
\begin{equation}
    \frac{DT}{T} = \frac{\sqrt{\langle \Delta^2(l) \rangle}}{\langle T^{\tau = 0}_{21}(z)\rangle} \ ,
    \label{eqn:angular_ps}
\end{equation}
where $\langle T^{\tau = 0}_{21}(z) \rangle$ is the mean brightness temperature of the simulation box at $z$.

We choose from \citet{holder2014} the two sets of observational data that are most constraining (quoting here 95\% confidence limits): a) an upper limit on $DT/T$ of $1.4\times 10^{-5}$ at 8.7 GHz and $\theta = 120''$ [corresponding to $l = 4040$ using $l \sim 2.35/\theta$ as in \citet{holder2014}] from ATCA \citep{Subrahmanyan2000}, and b) an upper limit on $DT/T$ of $6\times 10^{-5}$ at 4.86 GHz and $\theta = 60''$ (corresponding to $l = 8080$) from the VLA \citep{Fomalont1988}. We calculate the dimensionless angular fluctuation  (using eq. \ref{eqn:angular_ps}) at $l_1=4040$ and $l_2=8080$ and refer to it henceforth as $\left(\frac{DT}{T}\right)_1$ and $\left(\frac{DT}{T}\right)_2$, respectively. Defined in this way, we note that this ratio is independent of $f_{\rm{Radio}}$, since both the numerator and denominator are proportional to this parameter.

To compare to observations of clustering, we need to account for the fact that the CMB also contributes to the radio background at these frequencies. For a given $f_{\rm{Radio}}$, at $8.7$ GHz, the contribution of the high-redshift radio emission to the mean radio background is
\begin{equation}
    [T^{\tau = 0}_{21} (z)]_{8.7 \ \rm{GHz}} =\langle T^{f_{\rm{Radio}}=1}_{21}(z) \rangle f_{\rm{Radio}} \left[\frac{8700}{\nu(z)}\right]^{-2.7}  \ .
\end{equation}
Since the CMB dominates at these frequencies, the upper limits on $f_{\rm{Radio}}$ due to clustering, corresponding to the measured upper limits on the angular power spectrum at $l_1$ and $l_2$, are as follow:

\begin{equation}
    f_{\rm{Radio}}^{\rm{DT1}}(z) = \frac{1.4\times 10^{-5}\times2.725}{\left(\frac{DT}{T}\right)_1 \langle T^{f_{\rm{Radio}}=1}_{21}(z) \rangle \left[\frac{8700}{\nu(z)}\right]^{-2.7} } \ ,
\end{equation}

\begin{equation}
    f_{\rm{Radio}}^{\rm{DT2}}(z) = \frac{6\times 10^{-5}\times2.725}{\left(\frac{DT}{T}\right)_2 \langle T^{f_{\rm{Radio}}=1}_{21}(z) \rangle \left[\frac{4860}{\nu(z)}\right]^{-2.7} } \ .
\end{equation}

\section{Results}

\subsection{The constraints from radio clustering}

In order to study the constraints on high redshift radio clustering, we choose a model (which we refer to as our main case), where the seven astrophysical parameters are: star formation efficiency $f_{\star}=0.1$, $V_{\rm{c}}=16.5$ (corresponding to the minimum halo mass for star formation set by atomic cooling), a hard X-ray SED ($\alpha=1.5, E_{\rm{min}} = 1$ keV) with an efficiency ($f_{\rm{X}}=1$) corresponding to the typical observed value for low metallicity galaxies, an overall ionizing efficiency $\zeta=30$ and a maximum mean free path of ionizing photons ($R_{\rm{mfp}}$) of $30$ comoving Mpc. The constraints on $f_{\rm{Radio}}$ only depend on $f_{\star}$ and $V_{\rm{c}}$, plus a slight sensitivity at the low-redshift end to the reionization parameters (because of photoheating feedback due to reionization). 

\begin{figure}
    \centering
    \includegraphics[scale=0.5]{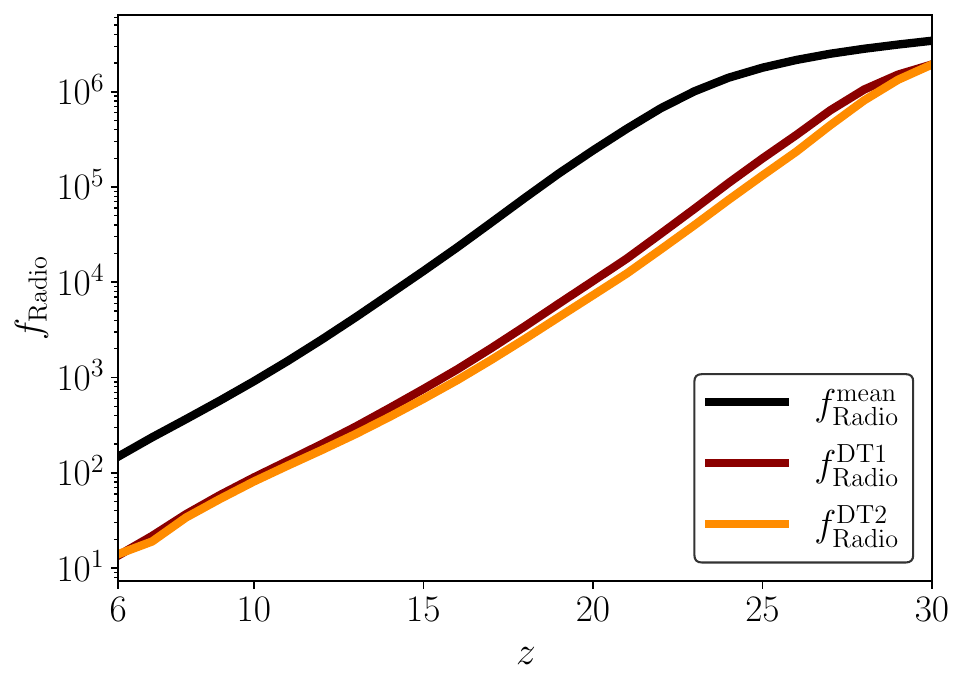}
    \caption{The radio efficiency $f_{\rm{Radio}}$ from our model as a function of $z$. We show (black line) the upper limit from the mean radio background  ($f^{\rm{mean}}_{\rm{Radio}}$) as a function of $z$, i.e., the highest value that is consistent with the maximum possible radio background from unresolved sources as measured by ARCADE-2. We also show the maximum values that are consistent with the observed upper limit on clustering at 8.7 GHz from ATCA (dark red line) and 4.86 GHz from VLA (orange line).}
    \label{fig:fRadio_z}
\end{figure}

As shown in the previous section, we find the upper limits on the radio efficiency parameter $f_{\rm{Radio}}$ at each redshift, from the observed upper limits on the mean radio background ($f^{\rm{mean}}_{\rm{Radio}}$) and from each of the two upper limits on clustering ($f_{\rm{Radio}}^{\rm{DT1}}$ and $f_{\rm{Radio}}^{\rm{DT2}}$). These values are shown in Fig.~\ref{fig:fRadio_z} as a  function of $z$. The values of
$f_{\rm{Radio}}^{\rm{DT1}}(z)$ and $f_{\rm{Radio}}^{\rm{DT2}}(z)$ are fairly similar, with $f_{\rm{Radio}}^{\rm{DT2}}(z)$ providing a slightly stricter limit than $f_{\rm{Radio}}^{\rm{DT1}}(z)$. We adopt the better limit (i.e., the lower of the two values) and refer to it as $f_{\rm{Radio}}^{\rm{DT}}(z)$. The values of $f_{\rm{Radio}}^{\rm{DT}}$ and $f^{\rm{mean}}_{\rm{Radio}}$ are also listed at various redshifts in Table~\ref{tab:fRadio_main_case}. 

\begin{table}
\centering
\begin{tabular}{lcccccc} 
\hline
\hline
Redshift   & 7  & 10 & 13 & 16 & 19 &  22    \\
\hline
$f^{\rm{DT}}_{\rm{Radio}}$  & 19.0 & 81.0  & 255 & 939 & 4320 & 22100 \\
\hline
$f^{\rm{mean}}_{\rm{Radio}}$ & 235 & 913 & 4280 & 23200 & 139000 & 669000 \\
\hline
\hline
\end{tabular}
\caption{Maximum $f_{\rm{Radio}}$ at various redshifts corresponding to the observed upper limits on the mean radio background ($f^{\rm{mean}}_{\rm{Radio}}$) or on fluctuations/clustering in the radio background ($f^{\rm{DT}}_{\rm{Radio}}$). The values correspond to the results shown in Fig.~\ref{fig:fRadio_z}. We assume our main astrophysical model.}\label{tab:fRadio_main_case}
\end{table}

Compared to the limit from the mean radio background ($f^{\rm{mean}}_{\rm{Radio}}$), which has been used in previous papers, the limit from clustering is lower (i.e., stronger) by between one and 1.5 orders of magnitude. Indeed, the relative factor by which the constraint improves is 12.4, 11.3, 16.8, 24.7, 32.2, and 30.3, at $z=7$, 10, 13, 16, 19, and 22, respectively. 

\subsection{Consequences for the 21-cm signal}

\begin{figure*}
\textbf{Main case}
\centering

\plottwo{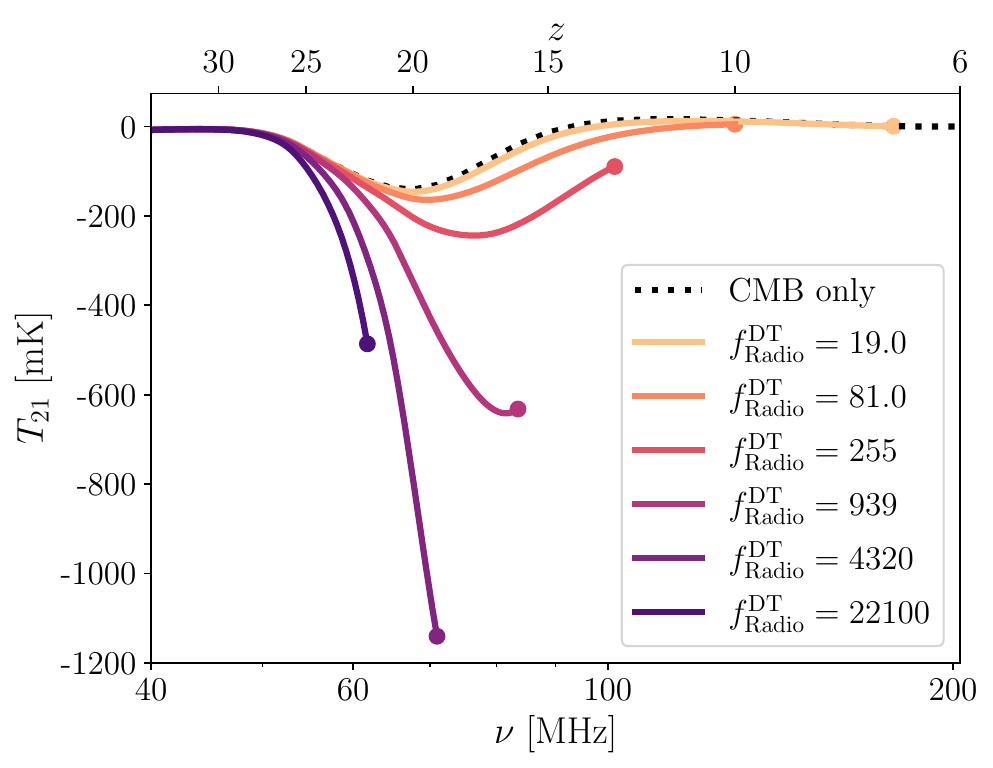}{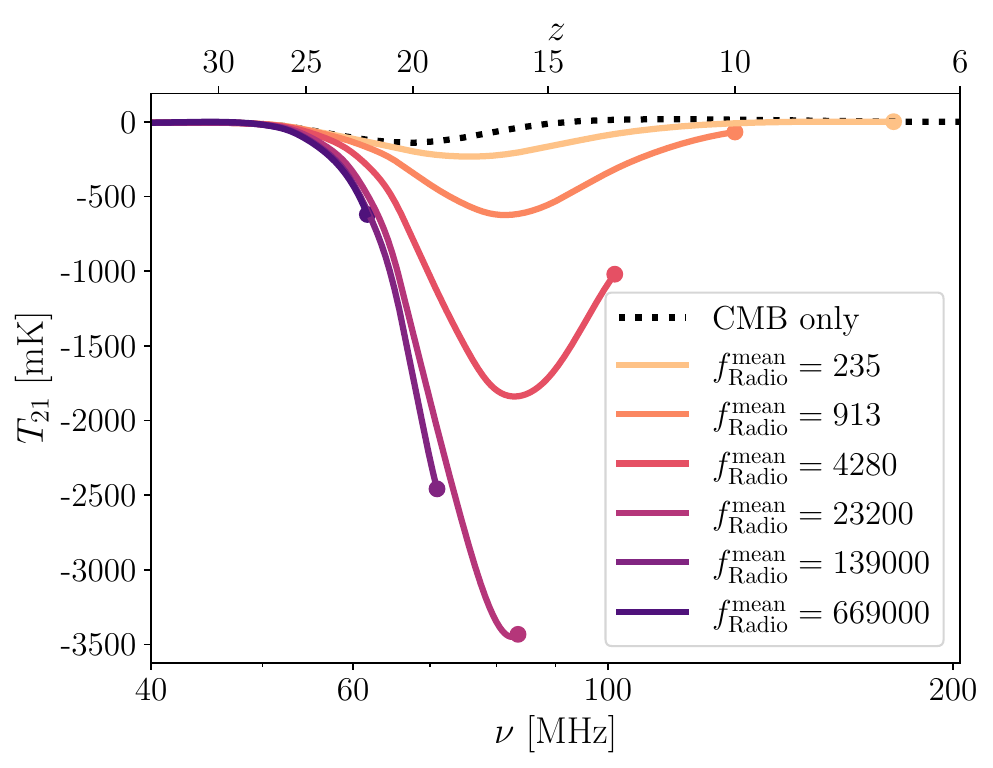}
\end{figure*}

\begin{figure*}
\textbf{Weak X-rays}
\centering

\plottwo{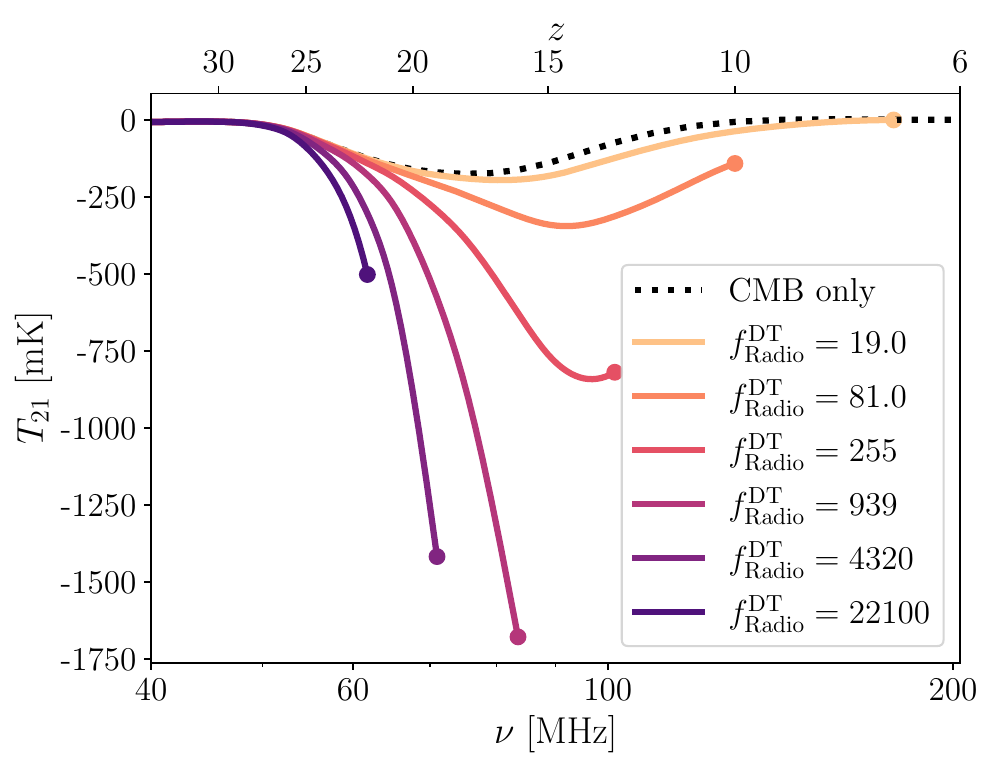}{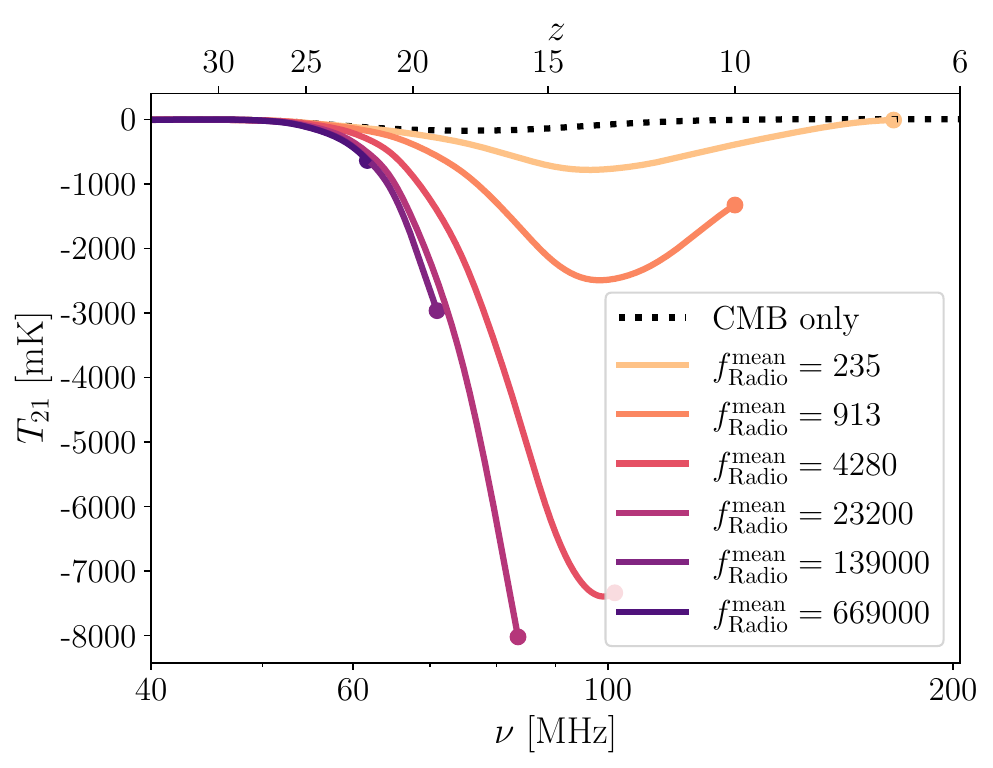}

\caption{The global 21-cm signal from cosmic dawn as a function of $\nu$ (or $z$ as the top x-axis) for various values of $f_{\rm{Radio}}$ from Table~\ref{tab:fRadio_main_case}. Each curve is cut off at the redshift at which the radio emission corresponding to $f_{\rm{Radio}}$ (also listed in the legend) must be truncated in order not to over-predict the observed extragalactic radio background. We show the maximum model as constrained by radio fluctuations (left panels) or the mean background (right panels), for our main astrophysical model (top panels) or the model with a low X-ray efficiency (bottom panels). We also show the case with no excess radio background ($f_{\rm{Radio}} = 0$), i.e., the CMB-background only case (dotted black line).}\label{fig:global_signal_main_add1}

\end{figure*}

In this section we explore the consequences of the radio clustering constraints for the 21-cm signal. Here we use eq.~(\ref{eqn:T21new}), i.e., the $\tau_{21} = 0$ terms have been subtracted to correspond to foreground removal in 21-cm observations (both global and  interferometer-based). We show our results for our main astrophysical model ("main case"), as given in the previous subsection, which represents a typical model. In order to have a better idea of the possible range of models, we also consider an additional model ("Weak X-rays"), in which the parameters are unchanged except that we assume a fairly low X-ray efficiency, $f_{\rm{X}} = 0.01$. This substantially enhances the signal since 21-cm absorption tends to give a stronger signal (both global and fluctuation) than 21-cm emission. Our experience with various model parameters suggests that this model represents something close to the maximum 21-cm signal at cosmic dawn. We comment further on the astrophysical parameters at the end of this subsection.

Fig.~\ref{fig:global_signal_main_add1} shows the maximum global 21-cm signal from cosmic dawn (and also the epoch of reionization) as a function of $\nu$ (or $z$ as the top x-axis). Each curve shows the prediction of a particular model with a fixed $f_{\rm{Radio}}$, corresponding to one of the values in Table~\ref{tab:fRadio_main_case} (and also indicated in the legend). Each curve starts from high redshift and terminates at the particular redshift at which the radio emission associated with that $f_{\rm{Radio}}$ value is the maximum allowed by observations; thus, each curve shows the full redshift range over which that model is consistent with observations of the extragalactic radio background. The top panels shows our main astrophysical model, while the bottom panels show the model with low X-ray efficiency. The left panels show models corresponding to the constraint from the fluctuations/clumping of the radio background, while the right panels show the models as constrained by the mean extragalactic radio background.

\begin{figure*}
\textbf{Main case}
\centering

\plottwo{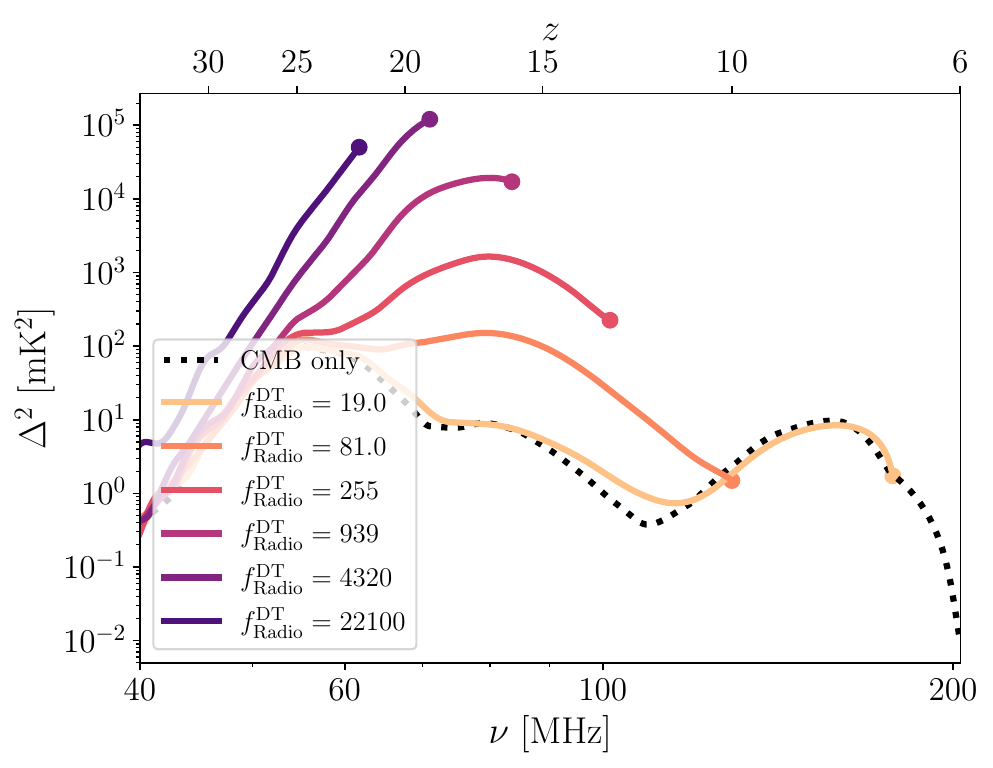}{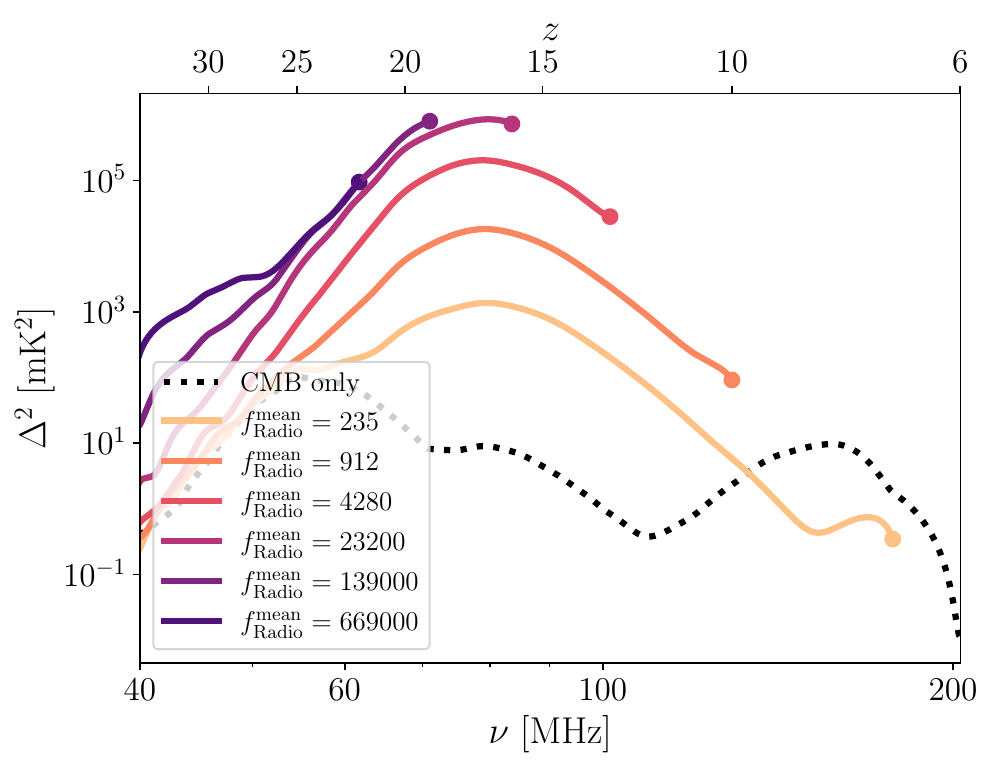}
\end{figure*}

\begin{figure*}
\textbf{Weak X-rays}
\centering

\plottwo{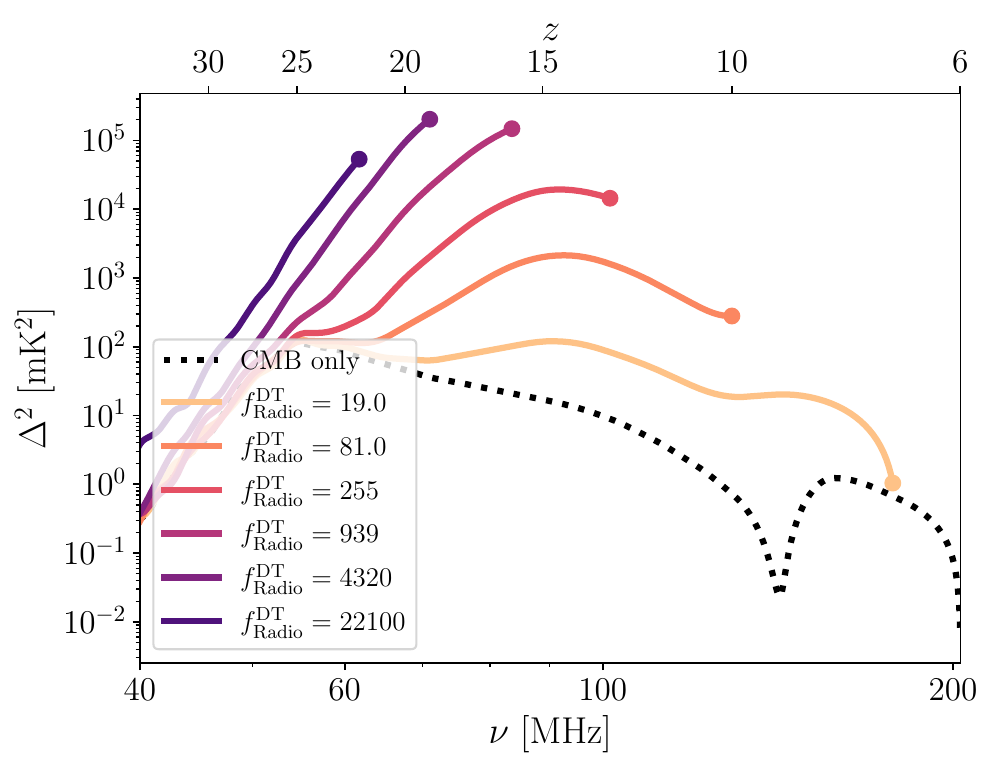}{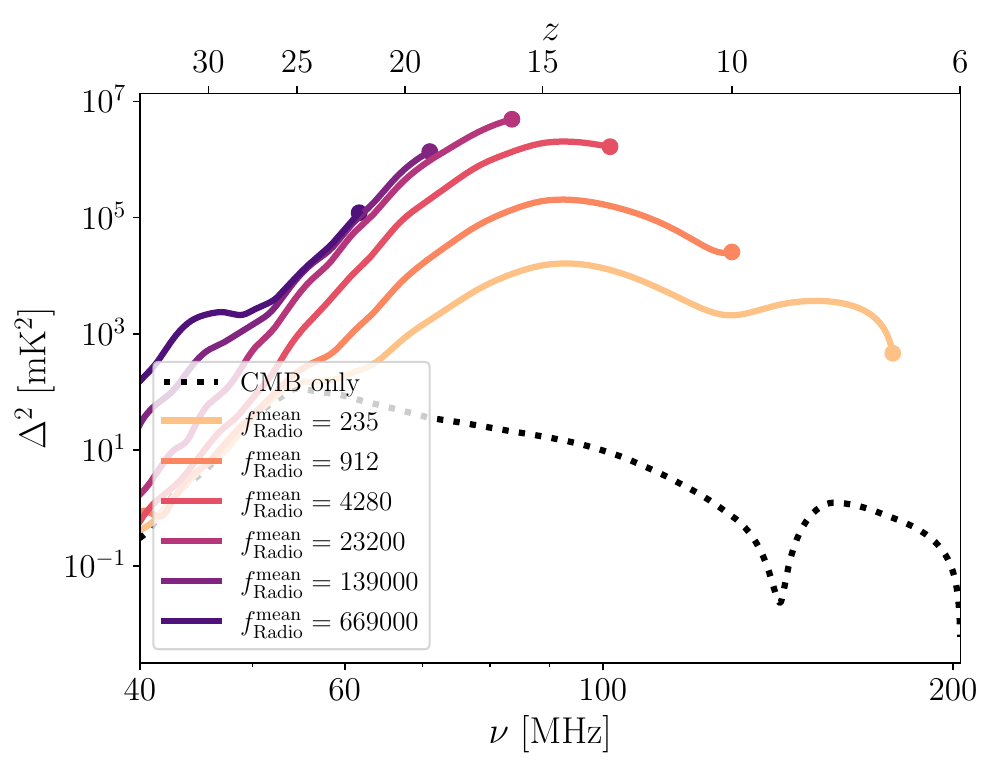}

\caption{Same as Fig.~\ref{fig:global_signal_main_add1}, except that here we show, instead of the global signal, the 21-cm power spectrum (in terms of the squared fluctuation) at $k=0.1$~Mpc$^{-1}$.}\label{fig:power_spectrum_main_add1}

\end{figure*}

For the main case, the maximum absorption allowed by the mean radio background is -3450~mK (at $z=16$), while including the radio clustering constraint reduces this by a factor of 3.03 to just -1140~mK (at $z=19$). The weak X-ray model gives a significantly stronger global signal, down to -8020~mK (at $z=16$) with only the mean background constraint, reduced $\times 4.77$ to -1680~mK (at $z=16$) with the clustering constraint. Thus, radio clustering is a much more important limit, but it still allows an absorption depth that can explain the EDGES detection and is more than 8 times larger than the same astrophysical model with no excess radio background (dotted black line). The clustering constraints are significantly stronger than the mean background constraint even at lower redshifts, down to the epoch of reionization, although there the relative change in the allowed $f_{\rm{Radio}}$ values is smaller, and also the actual allowed values of $f_{\rm{Radio}}$ are significantly smaller due to the larger galaxy populations at these redshifts. 

The other key observables in 21-cm cosmology are the statistical properties of the fluctuations, usually quantified by the 21-cm power spectrum. The 21-cm power spectra as a function of $\nu$ (or $z$ at the top x-axis) are shown in Fig.~\ref{fig:power_spectrum_main_add1}. As in Fig.~\ref{fig:global_signal_main_add1}, each curve shows a particular $f_{\rm{Radio}}$ value over the appropriate redshift range. Specifically we show the squared fluctuation (in mK$^2$) at $k=0.1$~Mpc$^{-1}$. For the main case, the maximum squared fluctuation allowed by the mean radio background is $8.57 \times 10^5$ mK$^2$ (at $z=19$), while including the radio clustering constraint reduces this by a factor of 7.08 to just $1.21 \times 10^5$~mK$^2$ (at $z=19$). As before, the weak X-ray model gives a significantly stronger signal, of $4.92 \times 10^6$~mK$^2$ (at $z=16$) with only the mean background constraint, reduced $\times 24.4$ to $2.02 \times 10^5$~mK$^2$ (at $z=19$) with the clustering constraint. Thus, the measured upper limits on radio fluctuations are significantly more constraining for the 21-cm power spectrum than the limits on the mean, overall radio background. 

The factor by which the limit improves on the squared fluctuation is roughly the square of the factor of the improvement on the absorption depth of the global signal; this is not exact since the 21-cm fluctuations induced by the fluctuating high-redshift radio background compete with other sources of 21-cm fluctuations (density, Lyman-$\alpha$ coupling, X-ray heating, reionization). Even with the reduced maximum 21-cm signal after inclusion of the radio clustering constraint, the 21-cm power spectrum at cosmic dawn redshifts ($13-20$) can still be two to four orders of magnitude larger than that expected from the same standard (CMB-background only) astrophysical scenario (dotted black line). This would increase by another order of magnitude if we only accounted for the observational constraint on the mean radio background. Since for a strong radio background, the main source of 21-cm fluctuations is initially radio fluctuations (plus those from Lyman-$\alpha$ coupling), lowering the X-ray efficiency does not affect much the 21-cm power spectrum at the highest redshifts, as can be seen from comparing the top and bottom panels of Fig.~\ref{fig:power_spectrum_main_add1} (the same is true in Fig.~\ref{fig:global_signal_main_add1}, although this is harder to see due to the linear $y$-axis that is standard for the global signal). However, a low X-ray efficiency boosts the signal, by up to 2 orders of magnitude compared to our main case, at lower redshifts ($< 15$) where heating fluctuations dominate.

We have illustrated the results for two particular sets of astrophysical parameters, a main case plus one with a rather low X-ray efficiency. While the particular constraints on $f_{\rm{Radio}}$ depend on the overall star formation history (as primarily determined by the parameters $f_{\star}$ and $V_{\rm{c}}$), the resulting maximum 21-cm signal at each redshift should be less sensitive to these parameters, since the overall radio emission (for a given $f_{\rm{Radio}}$) is proportional to the overall star formation rate. The 21-cm signal at various redshifts does depend in a complex manner on the various astrophysical parameters, but for the range of possible signals at cosmic dawn redshifts, the main parameter is the X-ray efficiency, which we have significantly varied. The 21-cm power spectrum (as opposed to the global signal) is sensitive to additional parameters, e.g., a higher galactic halo mass (corresponding to a higher $V_{\rm{c}}$) indicates rarer and more highly biased galaxies, and thus yields stronger 21-cm fluctuations (after fixing the overall star formation rate by raising $f_{\star}$). In future work, we plan to explore the radio clustering constraints over the full range of possible radio-excess astrophysical models. 

\section{Summary and Discussion}

We have explored new constraints on astrophysical models of high-redshift galaxies with a high efficiency of radio emission. Such models have been previously constrained by the overall extragalactic radio background as observed by ARCADE-2 and LWA-1, but we showed that a substantially stronger constraint comes from limiting the clustering of high redshift radio sources due to the observed upper limits on arcminute-scale anisotropy from the VLA at 4.9~GHz and ATCA at 8.7~GHz. To illustrate this, we used a semi-numerical simulation of a plausible astrophysical model, for either a vanilla model or one with a particularly strong 21-cm signal achieved due to a low efficiency of X-ray production.
We showed that the clustering constraints on the radio efficiency are stronger than those from the overall background intensity, by a factor that varies between 11 and 32, generally increasing with redshift over the range $z=7-22$. As a result, the predicted maximum depth of the global 21-cm signal is lowered by a factor of 4.8 (to 1700~mK) in the weak X-ray model, and a factor of 3.0 (to 1100~mK) in the vanilla model. The maximum 21-cm power spectrum peak at cosmic dawn is lowered by a factor of 24 (to $2.0\times 10^5$~mK$^2$) in the weak X-ray model, and a factor of 7.1 (to $1.2\times 10^5$~mK$^2$) in the vanilla model. 

We conclude that the observed clustering is currently the strongest direct constraint on such models, but strong early radio emission from galaxies remains viable for producing a strongly enhanced 21-cm signal from cosmic dawn. In particular, they can produce a global signal in the range of the possible EDGES detection, with an absorption depth that is 8 times larger than the same astrophysical model without a high radio efficiency. The corresponding 21-cm power spectrum peak during cosmic dawn is 3 orders of magnitude higher than for the same standard (CMB-background only) model. 

We note that in recent papers we and our collaborators have constrained excess radio models with current 21-cm data. Those constraints are more indirect than the radio background observations considered here, as the 21-cm signal is sensitive to various parameters and does not depend only on the radio emission from early galaxies. Also, in those papers the value of $f_{\rm{Radio}}$ was assumed to be fixed at all redshifts. In \citet{Bevins2024} we found a $1\sigma$ upper limit on $f_{\rm{Radio}}$ of 330, which is much weaker than the $2\sigma$ constraint we found here from clustering, especially given that \citet{Bevins2024} used data down to $z=8$. In \citet{Pochinda2023}, where we added the full radio line-of-sight fluctuations (as in the current paper) and also Population III galaxies (which were not included here), the upper limit derived for $f_{\rm{Radio}}$ was 52 at $1\sigma$ and 5400 at $2\sigma$. Thus, the constraints that we have derived from radio clustering (at $95\%$ confidence) should play a major role when added to current constraints. More generally, it is important to include the constraints from clustering when considering current and upcoming 21-cm experiments. 

\section{Acknowledgments}
SS and RB acknowledge the support of the Israel Science Foundation (grant No. 2359/20). AF was supported by the Royal Society University Research Fellowship.

%

\vspace{5mm}

\software{\texttt{Numpy} \citep{harris2020array}, \texttt{Scipy} \citep{2020SciPy-NMeth}, \texttt{matplotlib} \citep{Hunter:2007}}




\bibliography{sample631}{}
\bibliographystyle{aasjournal}



\end{document}